\documentclass[twocolumn,showpacs,pre,letterpaper]{revtex4-1}
\usepackage{hyperref}
\usepackage{graphicx}
\usepackage[utf8]{inputenc}
\usepackage{amsmath}

\def\de{\partial}
\begin{document}
%===============================================================================================================================
\title{Absence of thermalization for systems with long-range interactions
   coupled to a thermal bath}
%===============================================================================================================================
\author{Pierre de Buyl$^{1}$, Giovanni De Ninno$^{2,3}$, Duccio Fanelli$^{4,5}$, Cesare Nardini$^{4,6}$, Aurelio Patelli$^{4}$, 
Francesco Piazza$^{7}$, Yoshiyuki Y. Yamaguchi$^{8}$}
\affiliation
{\mbox{$^{1}$ Center for Nonlinear Phenomena and Complex Systems, Universit\'e libre de Bruxelles, B-1050, Brussels}\\
\mbox{$^{2}$ Laboratory of Quantum Optics, Nova Gorica University, Nova Gorica, Slovenia}\\
\mbox{$^{3}$ Sincrotrone Trieste, Trieste, Italy} \\
\mbox{$^{4}$ Dipartimento di Fisica e Astronomia, Universit\`a di Firenze and INFN}, \\
\mbox{Via Sansone 1, IT-50019 Sesto Fiorentino, Italy}
\mbox{$^{5}$ Dipartimento di Energetica ``Sergio Stecco'', Universit\`a di Firenze, via S. Marta 3, 50139 Firenze, Italy}\\
\mbox{$^{6}$ Laboratoire de Physique, ENS Lyon, Universit\'{e} de Lyon, CNRS, 46 All\'{e}e d'Italie, FR-69364 Lyon c\'{e}dex 07, 
France}\\
\mbox{$^{7}$ Universit\'{e} d'Orl\'eans and Centre de Biophysique Mol\'eculaire (CBM), Rue Charles Sadron, 45071 Orl\'eans, France}\\
\mbox{$^{8}$ Department of Applied Mathematics and Physics, Graduate School of Informatics, Kyoto University, Kyoto 606-8501, Japan} 
}
%===============================================================================================================================

%===============================================================================================================================
\begin{abstract}
We investigate the dynamics of a small long-range interacting system, in contact with a large long-range thermal bath. 
Our analysis reveals the existence of striking anomalies in the energy flux between the bath and the system. 
In particular, we find  that the evolution of the system is not influenced by the kinetic temperature of the bath, 
as opposed to what happens for short-range collisional systems. As a consequence, the system may get hotter 
also when its initial temperature is larger than the bath temperature. This observation is explained quantitatively in the framework of the collisionless Vlasov description of toy models with long-range interactions and shown to be valid whenever the
Vlasov picture applies, from cosmology to plasma physics.
\end{abstract}
%===============================================================================================================================
%
% PACS
%
% 05.20.-y Classical statistical mechanics
% 05.70.Ln Nonequilibrium and irreversible thermodynamics (see also 82.40.Bj Oscillations, 
%          chaos, and bifurcations in physical chemistry and chemical % physics)  
% 52.65.Ff Fokker-Planck and Vlasov equation
%
%
\maketitle
%===============================================================================================================================
%

\section{Introduction}

In the recent past, several theoretical and experimental studies have been devoted to exploring dynamical 
and thermodynamic properties of long-range interacting systems (LRIS's)~\cite{campa_et_al_phys_rep_2009}. In such systems, 
energy is not additive. This fact, together with a possible break of ergodicity, is at the origin of a 
large gallery of peculiar thermodynamic behaviors: the specific heat of LRIS's can be negative in the 
microcanonical ensemble~\cite{barre_et_al_prl_2001} and temperature jumps may appear at microcanonical first-order phase transitions. 
These systems also display remarkable non-equilibrium dynamical features. For example, it is well known that
under particular conditions isolated LRIS's may get trapped in long-lasting quasi-stationary states (QSS's), 
whose lifetime diverges with system size~\cite{yamaguchi_et_al_physica_a_2004,antoniazzi_et_al_pre_2007}. Importantly, when performing the limit
 $N\rightarrow \infty$ ($N$ being the number of particles), the system remains permanently confined in 
QSS's~\cite{antoniazzi_califano_prl,antoniazzi_et_al_prl_2007}. As a consequence, for large 
long-range interacting systems, QSS's are directly accessible through experiment~\cite{barre_et_al_pre_2004,bonifacio_de_salvo_1994,bachelard_et_al_jstat_2010}.

Until today, the large majority of studies aimed at elucidating the fundamental properties of LRIS's 
have been carried out on isolated systems, {\em i.e.} under the assumption that the system properties are 
not influenced by the external environment. 
However, recognizing whether a non-equilibrium QSS is stable to 
an external perturbation is of great importance~\cite{nardini_jstat_2012}, 
both from a theoretical and an experimental point of view. 
A related fundamental problem concerns the mechanism through which a LRIS exchanges energy with the surroundings. 
These questions epitomize the main motivation of the present work.    

The non-equilibrium dynamical properties of the LRIS's in contact with a thermal bath have been studied for the first 
time only recently~\cite{baldovin_prl_2006,baldovin_pre_2009,chavanis_noise_2011}. As a possible realization of thermal bath, these authors considered 
a large Hamiltonian system with nearest-neighbor interactions, coupled to a fraction of the spins in the system. 
They concluded that the coupling with the bath introduces a new time scale in the evolution of the system: 
the weaker the coupling strength, the longer the system remains trapped in a QSS 
before relaxing to equilibrium.

At variance with the above studies, we  investigate here the dynamics of a LRIS in long-range contact 
with an additional large system trapped in a QSS. 
This interaction scheme can be regarded as a more clear-sighted realization of 
a {\em thermal bath} for a LRIS, opening the way to 
applications in diverse fields such as cosmology and plasma physics. 
For example, one may think of the collisionless mixing between plasmas, or
the operation of magnetic fusion devices for energy production or the merging of globular clusters to a self-gravitating galaxy.
Furthermore, it is also tempting to speculate that our simple scheme could be somehow relevant 
for the self-consistent interaction between dark (the bath) and baryonic (the system) matter in the universe
(see, {\em i.e.} Ref.~\cite{springel_millenium_2005}).

\section{The long range thermal bath and the canonical QSS}

\indent As a reference case, we have selected the 
Hamiltonian Mean Field (HMF) model~\cite{antoni_ruffo_1995}, widely regarded as a prototype LRI system. 
The HMF Hamiltonian describes the one-dimensional motion of $N$ rotators coupled through a mean field cosine-like 
interaction,
\begin{equation}
\label{hamil}
H = \frac{1}{2} \sum_{j =1}^{N} p_{j}^2 + \frac{1}{2N}
\sum_{i,j=1}^{N} [ 1-\cos ( \theta_j - \theta_i)]
\end{equation}
where $\theta_j$ is the orientation of the $j$-th rotator and $p_j$ its conjugated momentum. 
To monitor the  evolution of the system, it is customary to introduce the magnetization $M$, an order parameter defined as
\begin{equation}
M = \frac{\left||\sum_i \mathbf{m}_i\right||}{N} \quad \mbox{where}
\quad \mathbf{m}_i = ( \cos \theta_i , \sin \theta_i) \label{l}.
\end{equation}
The infinite-range coupling between rotators is responsible for the
emergence of rather intriguing behaviors, including the existence of QSS's.
In a QSS the system displays non-Gaussian velocity distributions and it takes
values of $M$ different than those predicted
by equilibrium thermodynamics~\cite{antoniazzi_et_al_pre_2007,antoniazzi_califano_prl,benetti_prl_2012}. 

Rigorous mathematical results~\cite{braun_hepp_1977} indicate that in the limit $N\rightarrow \infty$ the discrete HMF dynamics 
reduces to the continuum Vlasov equation 
\begin{equation}
\label{vlasov}
\frac{\partial f}{\partial t}+p\frac{\partial f}{\partial \theta}-V^{\prime}(\theta) \frac{\partial f}{\partial p}=0,
\end{equation} 
where $f(\theta,p,t)$ is the microscopic one-particle distribution function, 
$V(\theta)\left[f\right]=-M_x[f]\cos(\theta)-M_y[f]\sin(\theta)$, 
$M_x[f]=\int_{-\pi}^{\pi}\int_{-\infty}^{\infty} f(\theta,p,t)\cos\theta \, d\theta \, dp$ 
and $M_y[f]=\int_{-\pi}^{\pi}\int_{-\infty}^{\infty} f(\theta,p,t)\sin\theta \,d\theta \,dp$. 
The specific energy $h[f]=\int \int (p^2/2)f(\theta,p,t)\,d\theta \,dp-\left(M_x^2+M_y^2-1\right)/2$ 
is a conserved quantity. The Vlasov equation defines the natural framework to address 
the puzzle of  QSS's emergence~\cite{yamaguchi_et_al_physica_a_2004,antoniazzi_califano_prl}. 
Specifically, QSS's are connected to the stable stationary solutions of the Vlasov equation. 
This observation suggests a statistical mechanics approach, inspired by the 
seminal work of Lynden-Bell~\cite{lynden-bell_1967}, to characterize analytically the QSS properties. 
Lynden-Bell's approach is based on the definition of a locally-averaged (``coarse-grained'') 
distribution, yielding an entropy functional defined from first-principle
statistical-mechanics prescriptions. By constrained maximization of such an entropy, 
one obtains closed analytical expressions for the single-particle distribution in the QSS 
regime~\cite{antoniazzi_califano_prl,antoniazzi_et_al_prl_2007}.  As a natural consequence, the QSS's can be equally 
interpreted as equilibrium configurations of the corresponding continuous description~\cite{staniscia_prl_2010}.
Hence, the QSS thermal bath that we consider here corresponds to a magnetized equilibrium solution of the 
underlying Vlasov equation~\eqref{vlasov}.

Let $f_B(\theta, p)$ be the normalized single-particle distribution that characterizes the QSS bath. 
Such a function is obtained as the stationary solution of the  Vlasov equation~\eqref{vlasov} 
corresponding to a water-bag initial distribution,
$f_0(p,\theta)=1/[4\Delta \theta_B \Delta p_B]$ for $\theta\in[-\Delta\theta_{B},\Delta \theta_{B}],
p\in[-\Delta p_{B},\Delta p_{B}]$ and zero elsewhere.
Note that the initial magnetization of the bath $(M_0)_B$ and its energy density 
$e_b$ can be expressed in terms of $\Delta \theta_B$ and $\Delta p_B$, as
$(M_0)_B=\sin \Delta \theta_B/\Delta \theta_B$ and $e_B=\Delta p_B^2/6+1/2(1-(M_0)_B^2)$. 
This in turn implies that the initial water-bag profile is uniquely determined by $(M_0)_B$ and $e_B$,
in agreement with the Lynden-Bell theory~\footnote{The Lynden-Bell theory 
provides a quantitatively correct description of macroscopic observables, such as the average  
QSS magnetization.  Alternative approaches accounting explicitly 
for non-ergodicity yield more accurate predictions~\cite{benetti_prl_2012}.}.

At this point, $t=0$ in our discussion, another HMF system with water-bag profile 
is injected and let evolve consistently with the bath. This system, $S$ in the following, 
is described in terms of its associated single-particle distribution  $f_S(\theta, p)$. 
Clearly the bath should be significantly larger than the system to which it is coupled. This can be accomplished through the following normalization condition
\begin{equation}
\label{norm}
\int f_S(\theta, p,t) \,d \theta \,d p  = 1 - \int f_B(\theta, p,t) \,d \theta \, d p = \epsilon 
\end{equation}  
where $\epsilon \ll 1$ sets the relative size of the two mutually interacting $S$ and $B$ HMF systems. 
We are interested in tracking the time evolution of the distribution  
$f(\theta,p,t) \equiv f_B(\theta, p,t) + f_S(\theta, p,t)$ under the constraint~\eqref{norm}. 
From the physical point of view, we are reproducing the microcanonical dynamics of one isolated HMF system ($S+B$), 
composed of two sub-systems supposed as distinguishable: the larger subsystem (the bath $B$) has already relaxed 
to its QSS equilibrium. The system $S$ is initially confined in an out-of-equilibrium configuration 
of the water-bag type. 
\indent To monitor the evolution of both subsystems, we follow the kinetic temperatures 
$T_{\alpha}(t) \equiv \Gamma_{\alpha} \int p^2 f_{\alpha}(p,\theta,t) \, d \theta \, d p$, 
with $\alpha=B,S$ and the corresponding magnetizations $M_{\alpha}$. Here, $\Gamma_{S}=1/\epsilon$ and $\Gamma_{B}=1/(1-\epsilon)$. 
We emphasize that $T_{\alpha}$ are average kinetic energies per particle and 
not true thermodynamic temperatures. In fact, our results highlight the crucial fact
that the appropriate definition of the true thermodynamic temperature associated 
with a QSS is not known.

A typical time evolution of these observables, obtained by numerical integration of the Vlasov equation~\eqref{vlasov},  is illustrated in Fig.~\ref{figure1}~\footnote{See Supplemental Material at \protect\url{http://link.aps.org/supplemental/10.1103/PhysRevE.87.042110}
for a movie depicting the time evolution of the bath and system phase portraits together with plots of 
the magnetization and the temperature.}.  
Before {\em injecting} the system ({\em i.e.}. at $t<0$), the bath is first prepared in a water-bag initial condition and then allowed to evolve towards a QSS. After the bath has relaxed well into its QSS (this is $t=0$), the interaction is switched on, meaning that a new HMF {\em combined system} is evolved, comprising bath and {\em system}.
After a short transient, the system reaches a quasi-equilibrium 
state where the mean value of the kinetic temperature is {\em different} from  the temperature of the bath.
In other words, the bath and the systems do not thermalize. Similarly, the two magnetizations converge to different values.
Importantly, we note that the specific values of temperature and magnetization attained
by the system spotlight a non-trivial interaction with the bath. $T_S$ and $M_S$ are indeed substantially different 
from the  values that the system would reach when evolved microcanonically from the same initial condition.
We obtain equivalent results upon simulating the discrete $N$-body dynamics~\eqref{hamil}. 
In this case, after a transient that gets progressively longer as the system size $N=N_S+N_B$ is increased, 
$\Delta T=T_B-T_S $ and  $\Delta M=M_B-M_S $ tend to zero. Thus, granularity causes thermalization, 
which is instead prevented in the continuum (Vlasov) limit.  
We term {\em canonical QSS's} the quasi-equilibrium configurations that the system explores when 
in long-range contact with a QSS thermal bath in the zero energy-flux regime. 

\begin{figure}
\resizebox{\columnwidth}{!}{\includegraphics{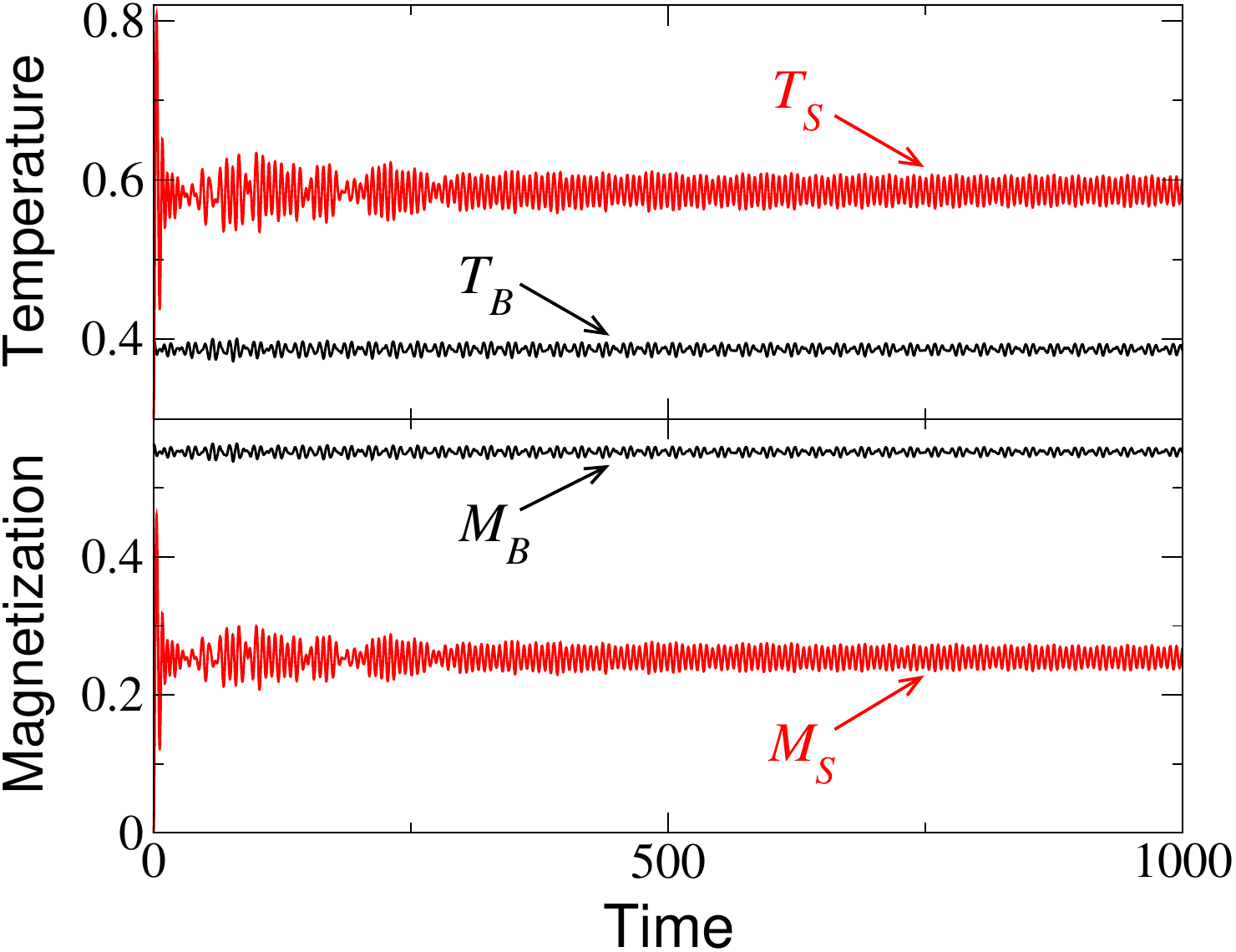}}
\caption{(Color online) Time evolution  of temperature  and magnetization. 
The bath QSS originates from a  water-bag with energy $0.54$ 
and initial magnetization $0.6$. The system is initially homogeneous in space ({\em i.e.}, zero magnetization) 
and its energy is set to 0.65. 
The coupling constant $\epsilon = 0.024$.
All quantities are dimensionless.}
\label{figure1}
\end{figure}
In the continuum limit, when the system is trapped in a canonical QSS, we find that 
the average energy flux between the bath and the system indeed vanishes, making the two subsystems 
by all means decoupled and thus preventing thermalization (see Appendix~\ref{app:flux} for a
more detailed analysis).  It is remarkable that a zero-flux steady state is reached for $T_B \ne T_S$ 
in the non-collisional continuum limit,  at variance with what is normally found in most collisional systems.

\section{The energy flux between the system and the bath}

\begin{figure}[t]
\resizebox{\columnwidth}{!}{\includegraphics{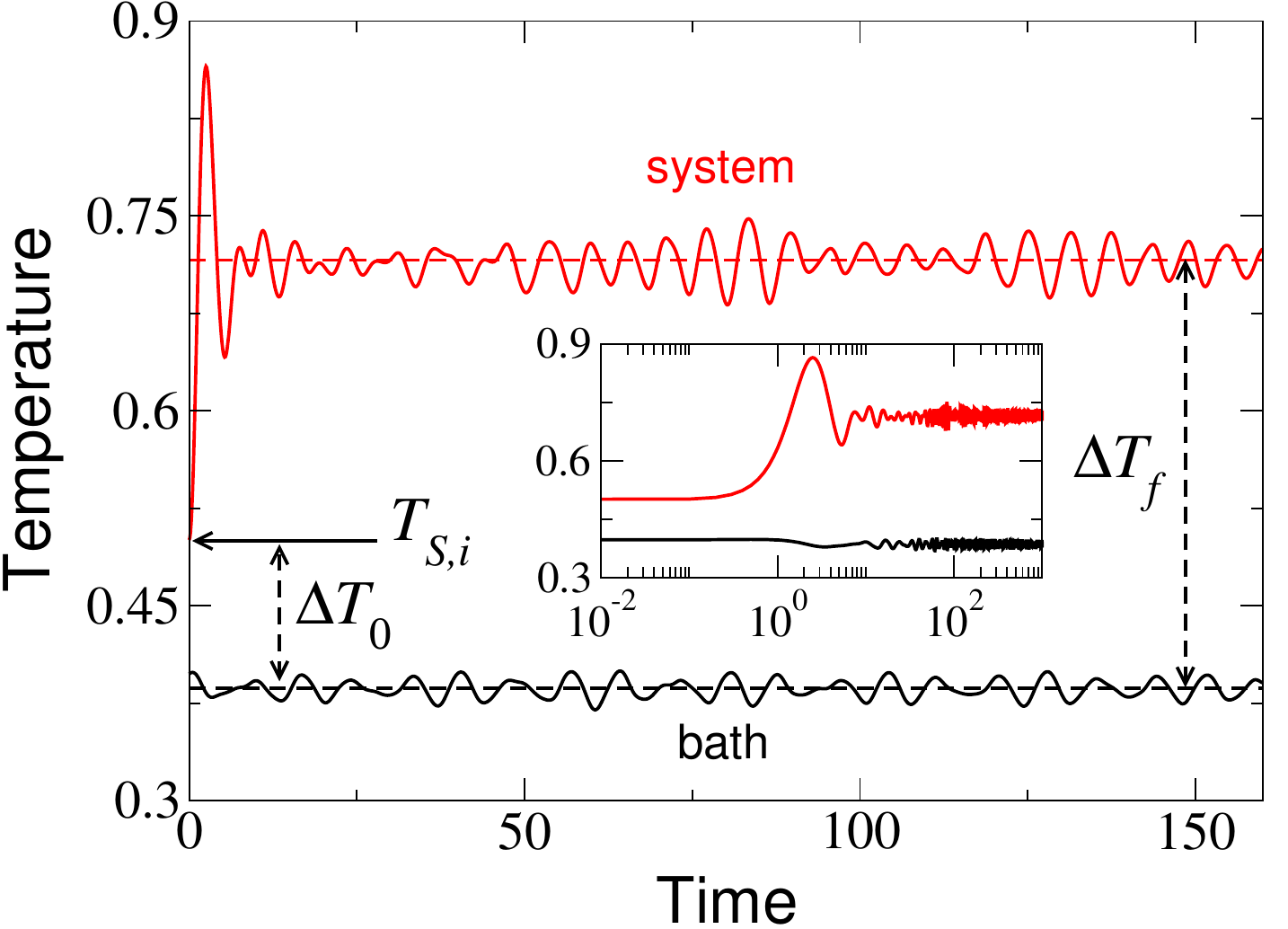}}
\caption{(Color online) Bath and system temperatures versus time. 
The system is initially space homogeneous (zero magnetization) and 
has energy $0.75$. Inset: same plot with logarithmic time scale. 
Other parameters are as in Fig.~\ref{figure1}. All quantities are dimensionless.}
\label{figure3}
\end{figure}

\indent Even more surprising is the behavior of the system during the ``violent relaxation''  stage
towards the canonical QSS, which is characterized by a net energy flux from the (cold) bath to the (hot) system.   
To better illustrate this observation, we plot  $T_B$ and $T_S$ versus 
time in Fig.~\ref{figure3}. Note that $T_S$ is larger than $T_B$ at $t=0$, the time of injection. 
As time progresses, the difference $\Delta T$ increases even further, resulting in an anomalous energy transfer 
from the bath to the system. In short, and counter-intuitively, the {\em hot} system gets {\em hotter} when placed
in contact with a large long-range QSS reservoir. This observation, although fighting intuition, does not violate any 
laws of physics, as the second law of thermodynamics is only expected to hold at thermal equilibrium.

Once the system has settled down in its canonical QSS at zero average energy flux, 
$\Delta T$ and $\Delta M$ are found to be different from zero. 
In order to pinpoint the relation between $\Delta T$ and $\Delta M$,
we performed a series of simulations for the same bath conditions as specified in the caption of Fig.~\ref{figure1}, 
and varying the initial energy of the system $S$. Different energies lead to distinct canonical QSS's, 
as it happens to isolated systems trapped in microcanonical QSS's. 
At first glance, it is tempting to speculate that canonical QSS's might originate from a  
net balance of two opposing thermodynamic forces, presumably related to $\Delta T$ and $\Delta M$. 
However, we find that the dynamical evolution of $S$ is {\em not} influenced 
by the temperature of the bath $T_B$, at least for $\epsilon \ll 1$,  
but only responds to its magnetization  $M_B$. 
Therefore, provided $M_B$ is kept fixed, $T_B$ can be set to an arbitrary value,  
without significantly altering the system dynamics. This is illustrated by the data collapse 
reported in Fig.~\ref{figure4}.

\section{Beyond the HMF model: a theoretical interpretation based on the Vlasov equation}

This striking observation is unintuitive as compared to the case 
of short-range systems. Even more interestingly, it is by no means restricted to the HMF.
In order to illustrate this fact, we note that in 
the Vlasov limit the distribution functions $f_{\alpha}(\theta,p,t)$  $(\alpha=B,S)$ obey to
\begin{equation}
\frac{\de f_{\alpha}}{\de t} + p\frac{\de f_{\alpha}}{\de \theta}
-V^{\, \prime}[f_{B}+f_{S}]\frac{\de f_{\alpha}}{\de p}=0
\label{eq:sysbath-ev}
\end{equation}
where $V(\theta)$ is a generic mean-field potential (the prime defining ordinary 
differentiation with respect to $\theta$), 
defined as
\begin{equation}
V[f]= \int f(\theta',p',t)v(\theta-\theta') \, d \theta'dp',
\end{equation}
$v(\theta-\theta')$ being the two-body potential.
Since the system/bath relative size $\epsilon \ll 1$,  we can treat it 
as a perturbative parameter, with $f_{S} \simeq \mathcal{O}(\epsilon)$ and
$f_{B} \simeq \mathcal{O}(1)$. Expanding eqs.~\eqref{eq:sysbath-ev} and 
keeping only terms that cause changes $\simeq \mathcal{O}(\epsilon)$ in the physical
observables, we are led to the two following coupled equations
\begin{equation}
\label{e:Vlasovcoup}
\frac{\de f_{\alpha}}{\de t} + p\frac{\de f_{\alpha}}{\de \theta}-V^{\, \prime}[f_{B}]\frac{\de f_{\alpha}}{\de p}=0
\qquad \alpha=B,S.
\end{equation}
The equation for the bath implies that this is frozen in its initial configuration, 
a stable equilibrium of the Vlasov equation, $f_{B}(q,p,t)=f_{B}(q,p,t=0)$ 
at all times~\footnote{A similar scenario is expected for baths at thermal equilibrium, which is also a
stable state of the Vlasov equation.}.
The equation for $f_{S}$ is the Liouville equation for a distribution of uncoupled 
particles moving in an external potential, $f_{B}$ being constant. 
These conclusions are utterly general and should apply to any 
physical system whose density is governed by the Vlasov equation. For the HMF model, thanks to its 
rotational invariance, one has with no loss of generality
\begin{equation}
\frac{\de f_{S}}{\de t} + p\frac{\de f_{S}}{\de \theta}-M_{B}\sin \theta\frac{\de f_{S}}{\de p}=0
\label{eq:system-evolution-HMF}
\end{equation}
which is simply the Liouville equation for a set of uncoupled pendula.
Hence,  the leading-order evolution of $f_{S}$ depends {\em only} on $M_{B}$ and {\em not} on $T_B$. 
As it is shown in Appendix~\ref{app:asympt}, $M_{B}$ sets the width of the 
resonance of the pendulum along $p$, which scales as $\sqrt{M_B}$. 
This implies that the temperature should be proportional to $M_B$, as can be also appreciated 
by dividing eq.~\eqref{eq:system-evolution-HMF} in the stationary state by $\sqrt{M_{B}}$.

\begin{figure}[t!]
\resizebox{\columnwidth}{!}{\includegraphics{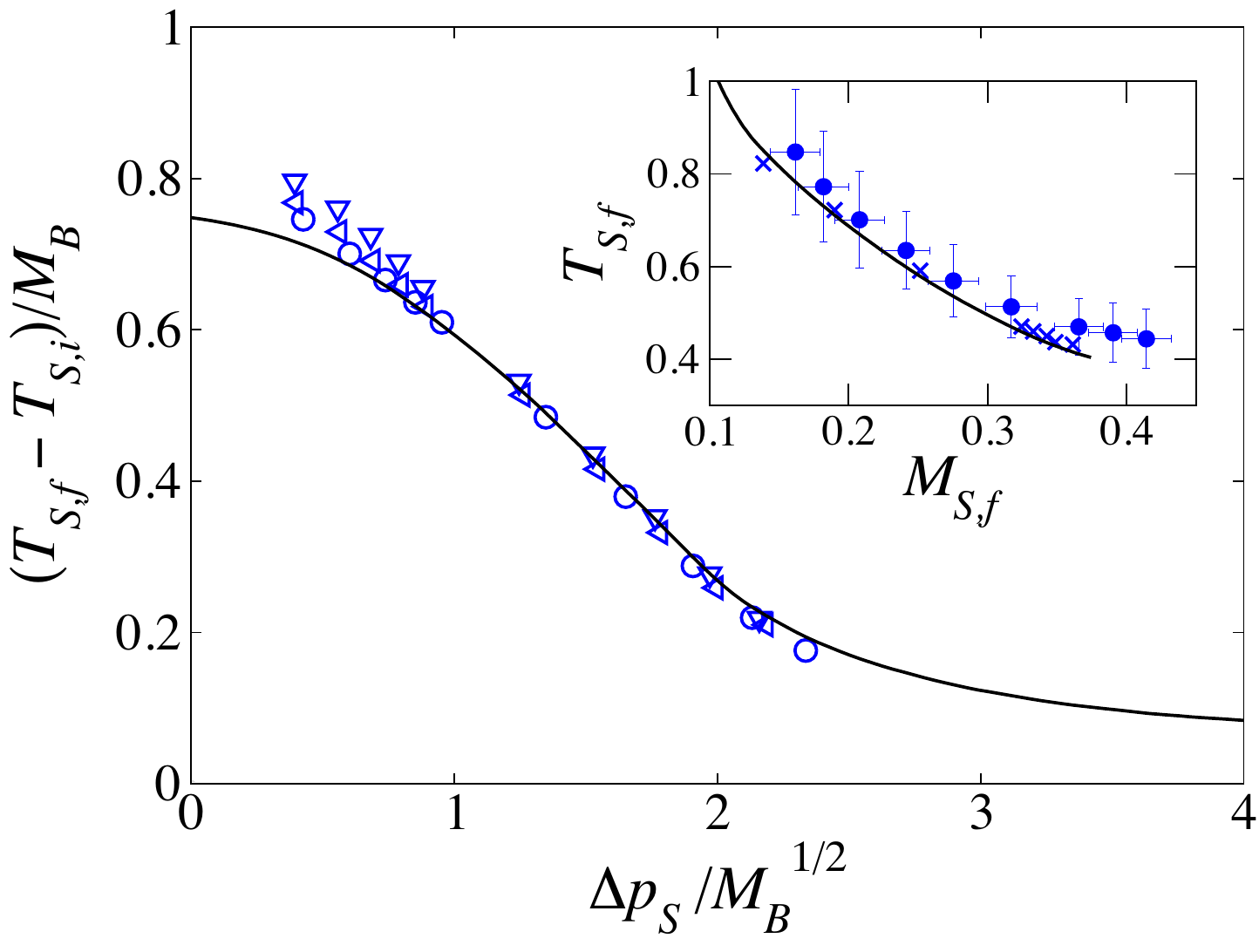}}
\caption{\label{figure4} (Color online) Difference between final and initial temperature of the system versus width of 
its initial water-bag, in reduced units. Data refer to different choices of the bath parameters and to different initial 
energies of the (initially homogeneous) system. Symbols: direct integration of eqs.~\eqref{eq:sysbath-ev}
for $M_{B}\in[0.1,0.55]$, $T_{B}\in [0.3,0.4]$. Solid line: numerical solution of eq.~\eqref{eq:system-evolution-HMF}. 
Inset: $T_{S,f}$ vs. $M_{S,f}$ for the same choice of parameters for the bath as in Fig.~\ref{figure1}. 
Circles: $N$-body simulations (average over $100$ independent realizations), $N_{B}=4\times10^{3},N_{S}=10^{2}$. 
Crosses: direct integration 
of the Vlasov equations. Solid line: integration of eq.~\eqref{eq:system-evolution-HMF}.
All quantities are dimensionless.}
\end{figure}

Consistently with the above scaling arguments, 
we plot in Fig.~\ref{figure4}  $(T_{S,f}- T_{S,i})/M_B$ as a function of the rescaled width of the initial
water-bag $\Delta p_{S}/\sqrt{M_B}$, for different values of the bath magnetization and temperature. 
The data refer to direct integration of the (constrained) Vlasov 
equations~\eqref{eq:sysbath-ev} and of eqs.~\eqref{eq:system-evolution-HMF}. In all cases, the data collapse 
nicely on a single master curve, which confirms the validity of our reasoning. 
An analytical calculation of $(T_{S,f}- T_{S,i})/M_B$ for $\Delta p_{S}=0$ yields 
$(T_{S,f}- T_{S,i})/M_B \approx 0.751$, in excellent agreement with the result of 
direct integration of eq.~\eqref{eq:system-evolution-HMF} (see Appendix~\ref{app:asympt} and also Ref.~\cite{de_buyl_et_al_self-consistent_pre_2011}).
The inset further shows that $N$-body simulations agree with all results obtained in the continuum limit.

\section{Conclusions}

Summarizing, we have proposed an implementation of long-range QSS bath. 
We showed that a small system in {\em true} long-range contact with a 
large, long-range reservoir reaches a zero-flux steady state, that we term {\em canonical} 
quasi-stationary state. These are stationary states of the system-bath coupled Vlasov equations, 
but quasi-stationary solutions of the associated $N$-body problem.
Remarkably, in the explored range of parameters, we find that 
hotter-than-bath systems become hotter in canonical QSS's. In the context of the HMF model,
based on simple scaling arguments, we have unveiled how the system anomalously increases its kinetic temperature 
as the fraction of its particles trapped in the resonance set by the bath magnetization 
gain energy.  The kinetic energy gain is proportional to the value of $M_B$ and independent
of the bath temperature at the leading order in $\epsilon$. We stress here, that this observation does not violate any fundamental laws of physics. Indeed, the average kinetic energy of the system does not coincide with its thermodynamic temperature. In this respect, our work raises the following central, yet unanswered, question: what is the correct thermodynamic measure of temperature for a system frozen in a QSS? Notice that in the present work, the energy of the 
thermal bath was chosen to lie in the part of the (microcanonical) phase diagram corresponding 
to a magnetized QSS. As regards the system, we considered initial energies leading to both magnetized and non magnetized
(microcanonical) QSS's. 

In conclusion, and based on the theoretical analysis that we have carried out, we argue that the 
results illustrated in this paper are general and extend beyond the HMF case-study, 
whenever the collisionless Vlasov picture is a good description of the dynamics.

\acknowledgments

Dicussions during the workshop ``Equilibrium and out-of-equilibrium
properties of systems with long-range interactions'' held at the Centre
Blaise Pascal, ENS-Lyon in August 2012 are acknowledged.
F.P., D.F. and G.D.N. would like to thank D. Hane for insightful discussions.

\appendix
%%%%%%%%%%%%%%%%%%%%%%%%%%%%%%%%%%%%%%%%%%%%%%%%%%%%%%%%%%%%%%%%%%%%%%%%%%%%%%%%%%%%%%%%%%%%%%%%%%%%%%%%%%%%%%%%%
%
%  APPENDIX A
%
\section{The energy flux}
\label{app:flux}

\noindent The energy flux from the bath $B$ to the system $S$ is defined as 
$\Phi_{B \rightarrow S} = - dE_B / dt$, where $E_B$ is the total energy of $B$. 
In order to derive an explicit expression for $\Phi_{B \rightarrow S}$, we start 
by calculating $\phi_j (t)$, the rate of energy loss of the $j$-the particle. 
Denoting by $h_j$ its energy, we have
\begin{equation}
  \label{phi_j}
  \varphi_j (t) \equiv -\frac{dh_j (t)}{dt} =  -\frac{1}{2}\,p_j\,{\bf M} \cdot {\bf m}^{\perp}_j  + 
  \frac{1}{2}  \frac{d \bf M}{d t} \cdot  {\bf m}_{j}
\end{equation}
where ${\bf m}_j \equiv ( \cos\theta_j , \sin\theta_j)$ and
${\bf m}^{\perp}_j \equiv ( -\sin\theta_j , \cos\theta_j)$, ${\bf M}$ being the global magnetization
\begin{equation}
 {\bf M}  =  (M_x,M_y) = \frac{1}{N}\sum_{j=1}^{N} {\bf m}_j
\end{equation}
Here $N$ is the total number of particles, {\em i.e.} the sum of those belonging to the bath, 
$N_B$, and those in the system, $N_S$. 
Summing over all particles belonging to the bath in eq.~\eqref{phi_j}, 
one eventually obtains
\begin{equation}
  \Phi_{{B \rightarrow S}} = \sum_{j\in B} \phi_j = -\frac{1}{2}\sum_{j\in B} p_j \, {\bf M} \cdot {\bf m}^{\perp}_j
  + \frac{N_B}{2} \frac{d \bf M}{d t} \cdot {\bf M}_B  ~.
\end{equation}
where ${\bf M}_B = \sum_{j\in B} {\bf m_j} / N_B$ is the magnetization of the bath and
the time derivative of ${\bf M}$ reads 
\begin{equation}
  \frac{d \bf M}{d t} = \frac{1}{N} \sum_{j=1}^N p_j\,{\bf m}^{\perp}_j 
\end{equation}
In the continuum limit the sums are replaced by integrals
\begin{equation}
  \Phi_{B \rightarrow S} = -\frac{1}{2} \int  p \, {\bf M} \cdot {\bf m}^{\perp}(\theta) f_{B}(p,\theta)\, d\theta \, dp 
                              + \frac{1}{2}  \frac{d \bf M}{d t} \cdot {\bf M}_B 
\end{equation}
where
\begin{equation}
{\bf M}_B  = \int f_B {\bf m}(\theta) \, d \theta \, d p  \quad 
{\bf M}    = \int \left(f_B + f_S\right) {\bf m}(\theta) \, d \theta \, d p
\end{equation}
with ${\bf m}(\theta) = (\cos \theta,\sin \theta)$, 
${\bf m}^{\perp}(\theta) = ( -\sin\theta, \cos\theta)$.

According to the adopted sign convention,  $\Phi_{B \rightarrow S}(t)$ is positive if 
the bath $B$ cedes energy to the system. In Fig.~\ref{figure2} the instantaneous 
energy flux (upper panel) is plotted versus time for a typical realization of the Vlasov dynamics. 
After an initial transient, $\varphi_{B \rightarrow S}(t)$ oscillates around zero, implying that the bath $B$ and the 
system $S$ have established a zero-average-flux dynamical equilibrium. This condition corresponds to the emergence of the  
canonical QSS. 
Furthermore, the net energy flux is positive, a fact that can be appreciated by 
looking at the evolution of the cumulated flux (see lower panel of Fig.~\ref{figure2}). 
This implies a net transfer of energy from the bath to the system. 

\begin{figure}[ht!]
\begin{center}
\includegraphics[width=\columnwidth]{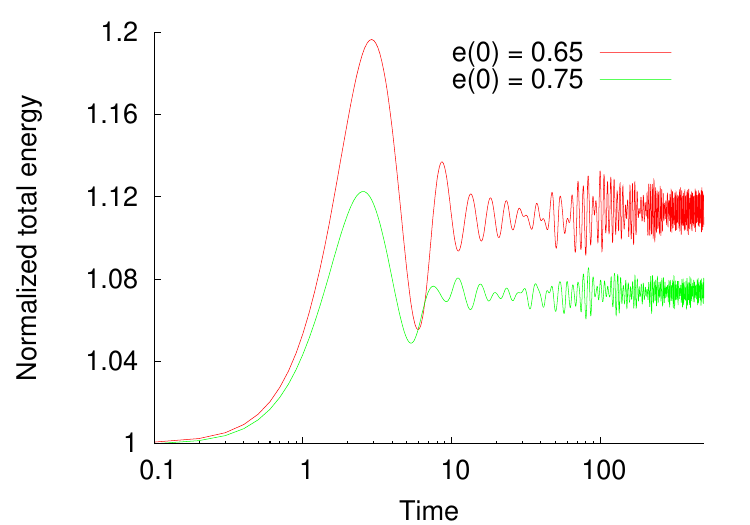}
\caption{Total energy of the system in contact with a large reservoir in 
a QSS for two different initial energies, normalized by the $t=0$ energy $e(0)$. 
The time $t=0$ marks the moment 
where the system and the bath are put in contact. Following the injection, the combined
(system$+$bath) ensemble is isolated, and hence its 
energy $E_{t}$ stays constant. Other parameters are: $M_{0}=0.6$
(initial magnetization of the bath), $E_{t}=0.547$ ($e(0) = 0.65$),
$E_{t}=0.55$ ($e(0) = 0.75$). All quantities are dimensionless.}
\label{f:obeccatiquesta}
\end{center}
\end{figure}

\begin{figure}[t]
\resizebox{\columnwidth}{!}{\includegraphics{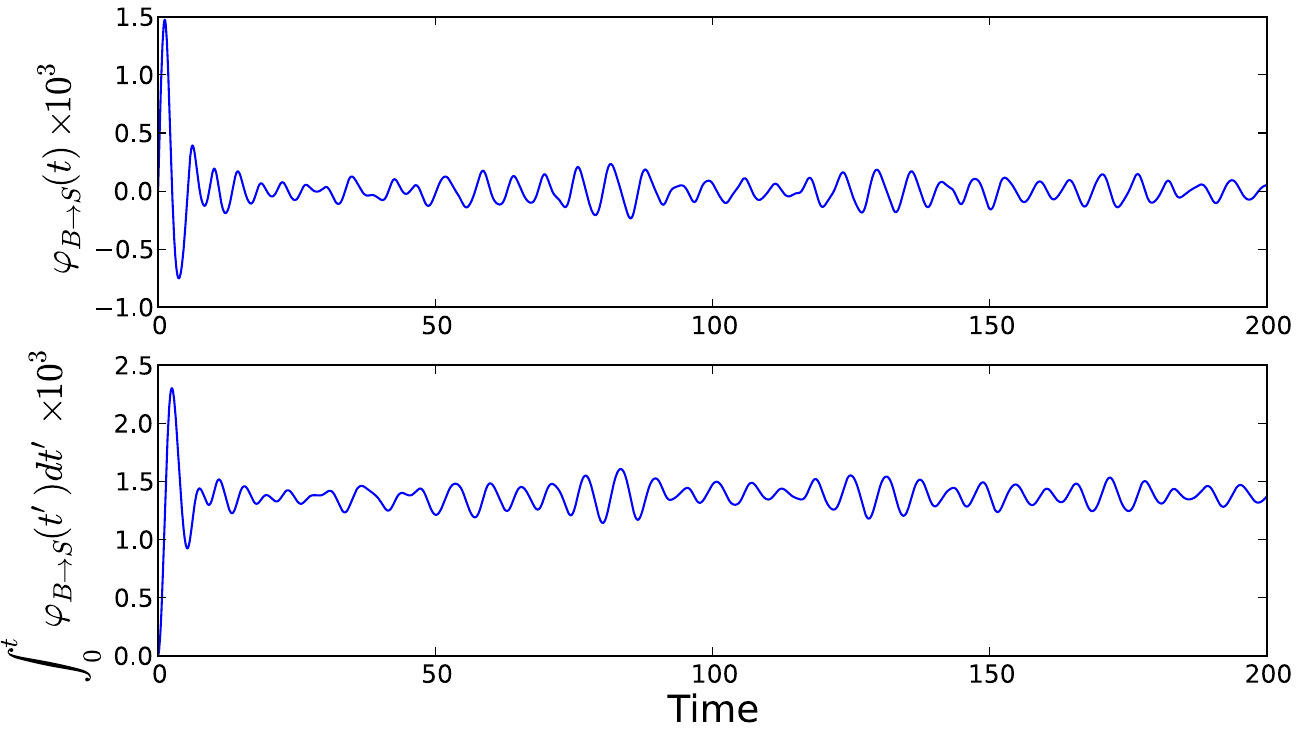}}
\caption{(Color online) Time evolution of the instantaneous (top) and cumulated (bottom)
bath-to-system energy flux. The system is initially space homogeneous and 
has energy $0.75$. Other parameters are as in Fig.~1 in the paper.
All quantities are dimensionless.}
\label{figure2}
\end{figure}

We stress that the system gets hotter as its {\em total} energy increases after putting it in contact 
with the bath. The total energy of the system increases when it is put in contact with the bath, as it is
clearly proved by looking at the energy flux (the time derivative of the total energy) versus
time in Fig.~\ref{figure2}. The cumulated flux is 
positive, which, according to our conventions, attests to a flow 
of total energy from the bath to the system.
In order to make this point even more clear, we show in Fig.~\ref{f:obeccatiquesta}
the total energy of the system versus time from the moment of the {\em injection}.
The total energy of system and bath stays constant, while there is a clear 
flux of {\em total} energy from the bath to the tiny system, which is left permanently 
hotter as a result. 

%%%%%%%%%%%%%%%%%%%%%%%%%%%%%%%%%%%%%%%%%%%%%%%%%%%%%%%%%%%%%%%%%%%%%%%%%%%%%%%%%%%%%%%%%%%%
%%%%%%%%%%%%%%%%%%%%%%%%%%%%%%%%%%%%%%%%%%%%%%%%%%%%%%%%%%%%%%%%%%%%%%%%%%%%%%%%%%%%%%%%%%%%

%==========================================================================================================================
%==========================================================================================================================

%%%%%%%%%%%%%%%%%%%%%%%%%%%%%%%%%%%%%%%%%%%%%%%%%%%%%%%%%%%%%%%%%%%%%%%%%%%%%%%%%%%%%%%%%%%%%%%%%%%%%%%%%%%%%%%%%
%
%  APPENDIX B
%
\section{On the analytic estimate of the asymptotic temperature $T_{S,f}$.}
\label{app:asympt}

\noindent The phase space of the pendulum is foliated in trajectories with constant energy
\begin{equation}
\label{separatrix}
  e = \frac{p^2}{2} - M_B \cos\theta
\end{equation}
hence, $p(\theta)= \sqrt{2} \left[ e + M_B \cos\theta \right]^{1/2}$. We want
to discuss an analytic estimate of the quantity
$(T_{S,f}- T_{S,i})/M_B$ for $\Delta p_{S}=0$ and for an initial homogeneous 
system, $M_S(t=0)=0$. This calculation has the merit of enabling one to gain insight 
into the nature of the canonical QSS and further clarify the scaling adopted in Fig.~4. 
This analysis can be extended to cover the case  $\Delta p_{S} \ne 0$, and also $M_S(t=0)\neq0$, 
a generalization to which we shall return in a separate contribution. 

We note that  $T_{S,i}=0$ for $\Delta p_{S}=0$. To evaluate $T_{S,f}$, 
we first consider the average kinetic temperature 
of the particles which are assigned a given energy $e$. In formulae
\begin{equation}
\label{average}
\langle p^2 \rangle_e =  \frac{1}{T(e)} \int_0^{T(e)} \dot{\theta}^{2} \, dt
\end{equation}
where  $\langle \cdot \rangle_e$ indicates a time average over one period 
\begin{equation}
\label{period}
T(e) =   \frac{4}{\sqrt{M_B}} K\left( \frac{e+M_B}{2 M_B} \right)
\end{equation}
$K(\cdot)$ being the complete elliptic integral of the first kind.
Expression~\eqref{average} takes the equivalent form
\begin{equation}
\label{average1}
\langle p^2 \rangle_e =  \frac{2}{T(e)} \int_{-\bar{\theta}(e)}^{\bar{\theta}(e)} p(\theta) \, d \theta 
\end{equation}
where $\bar{\theta}(e)=\cos^{-1}(-e/M_B)$ is the angle of inversion of the selected (closed) trajectory. 
By performing the integral one eventually gets
\begin{equation}
\label{average2}
\frac{\langle p^2 \rangle_e}{M_{B}} = 
\frac{2\sqrt{2(M_B+e)}} {\sqrt{M_{B}}K\left(\frac{e+M_B}{2 M_B} \right)} E\left( \frac{\bar{\theta}(e)}{2},
\frac{2 M_B}{e+M_B} \right)
\end{equation}
where $E\left( \cdot, \cdot \right)$ is the incomplete elliptic integral of the second kind. 
The final temperature of the system can now be evaluated as
\begin{equation}
\label{T_f}
T_{S,f} \equiv \langle p^2 \rangle = \int_{-M_B}^{M_B} \langle p^2 \rangle_e \rho(e)  \, de
\end{equation}
where $\rho(e)$ is the density of states of the system, which is univocally fixed by the initial condition. 
The integral in eq.~\eqref{T_f} extends from ${-M_B}$ to ${M_B}$, {\em i.e.} the energies that identify the 
separatrix of the pendulum.  In fact, the system is trapped inside the separatrix $|e|=M_B$, 
given the specific condition selected here ($\Delta p_{S}=0$, hence no particle lies outside 
the resonance at $t=0$). Recalling eq.~\eqref{separatrix}, the distribution 
$\rho(e)$ can be calculated easily, as
\begin{equation}
\label{rho}
\rho(e) =  \frac{1}{\pi} \left| \frac{d e}{d \theta} \right|^{-1} =
\frac{1}{\pi} \frac{1}{\sqrt{M_B^2-e^2}} 
\end{equation}
Plugging eq.~\eqref{rho} into eq.\eqref{T_f} and recalling eq.~\eqref{average2},
one eventually obtains
\begin{equation}
\label{T_f1}
\frac{T_{S,f}}{M_B} = \sqrt{\pi} \int_{-1}^{1}  
\frac{E\left( \cos^{-1}(-y)/2, \frac{2}{1+y} \right)}{ K\left( \frac{1+y}{2} \right)} \frac{dy}{\sqrt{1-y^2}}
\end{equation}
Numerical integration gives ${T_{S,f}}/{M_B} \approx 0.751$, in excellent agreement 
with the data reported in Fig.~3. In the general case ($\Delta p_{S} \neq 0$), $e \propto \Delta p_{S}^2$.
The scaling suggested by eq.~\eqref{T_f1} implies $\Delta p/ \sqrt{M_B}$, which in turn explains the origin 
of the reduced variables used in Fig.~\ref{figure4}.

\bibliography{/home/pierre/code/bibfile/pdebuyl}

%==================================================================================================================== 
\end{document}